\documentclass{webofc}
\usepackage[varg]{txfonts}   
\usepackage{siunitx}
\DeclareSIUnit\barn{b}
\DeclareSIUnit\clight{\text{\ensuremath{c}}}
\usepackage{hyperref}
\usepackage{url}
\usepackage{graphicx}
\usepackage{subcaption}
\usepackage{hepnicenames}
\usepackage{xpatch}
\makeatletter
\xpatchcmd\@HepConStyle
{\edef\@upcode{\updefault}}
{\ifdefined\shapedefault\edef\@upcode{\shapedefault}\else\edef\@upcode{\updefault}\fi}
{}{}
\makeatother
\newcommand{\pt}{\ensuremath{p_\text{T}^{}}\xspace}

\hypersetup{colorlinks=true,citecolor=blue,urlcolor=blue,linkcolor=blue}

\begin{document}
\title{New probes of nuclear gluon dynamics through photoproduction of charm in inelastic ultra-peripheral Pb–Pb collisions with ALICE}

\author{\firstname{Sigurd} \lastname{Nese}\inst{1}\fnsep\thanks{\email{sinese@cern.ch}}
        for the ALICE Collaboration
}

\institute{University of Oslo
          }

          \abstract{\unboldmath In an ultra-peripheral collision, a photon can interact with a gluon in the target nucleus and produce a pair of charm quarks, while the target nucleus breaks up (inelastic scattering). These charm quarks then fragment and are observed as open charm hadrons or vector mesons. This process has been used in e-p collisions to set stringent limits on the proton gluon distribution at low-$x$. The current measurements can provide similar constraints on the much less known nuclear gluon distributions. ALICE has measured the transverse momentum distribution of inelastically photoproduced $\text{D}^0$ at midrapidity, in Pb--Pb collisions at $\sqrt{s_\text{NN}}=\qty{5.36}{\TeV}$. The distribution is measured down to $\pt=0$ for the first time. The results are compared to model calculations.}
\maketitle
\section{Introduction}
\label{intro}
In ultra-peripheral collisions (UPCs), the impact parameter is larger than the sum of the nuclear radii. Since the nuclei do not overlap, hadronic interactions are strongly suppressed and the nuclei mainly interact electromagnetically. Photon-photon and photo-nuclear reactions are induced by the highly Lorentz-contracted electromagnetic fields of the colliding nuclei, which can be described as a flux of quasi-real photons. The dominant production mechanism for \HepProcess{\Pcharm\APcharm} pairs in UPCs is through photon-gluon fusion, \HepProcess{\Pphoton+\Pgluon\to\Pcharm\APcharm}. Photoproduction of charm in Pb--Pb UPCs at the Large Hadron Collider (LHC) can therefore be used to probe the nuclear gluon distribution. With low-\pt charm hadrons reaching Bjorken-$x$ values of $\sim 10^{-4}$, this process is sensitive to gluon saturation and nuclear shadowing. In Run 3 of the LHC, ALICE records large numbers of inelastic UPC events, owing to the use of continuous readout and a 50 kHz Pb--Pb interaction rate~\cite{ALICE:2023udb}.

\section{Experimental methods for the measurement of photoproduced \texorpdfstring{\PDzero}{D0}}
\label{sec-1}
In UPC photoproduction, a photon from the emitter nucleus may interact inelastically with the target nucleus, resulting in particle production. Due to the relatively low energy of the photon compared to the target, the production of particles is shifted in the target-going direction. Such events are therefore characterized by large rapidity gaps in the direction of the photon emitting nucleus, i.~e. intervals in rapidity where no particles are produced.

In ALICE, rapidity gaps are identified by the Fast Interaction Trigger system (FIT)~\cite{Trzaska:2017reu}, which sits on either side of the interaction point at large pseudorapidities. In the current analysis, the veto selection was done using the FT0-A and FT0-C, which are two of the detector arrays which comprise the FIT. They cover pseudorapidities of $3.5<\eta<4.9$ and $-3.3<\eta<-2.1$, respectively.  Inelastic photonuclear events are selected by requiring exactly one of the two FT0 arrays to have activity below a noise threshold. The side with low activity is identified as the photon-going direction, while the opposite side is identified as the target-going direction. The two event classes are denoted as $+\eta$ gap events and $-\eta$ gap events, depending on which side of the interaction point the rapidity gap was observed.

\begin{figure}
  \begin{center}
    \includegraphics[width=0.45\textwidth]{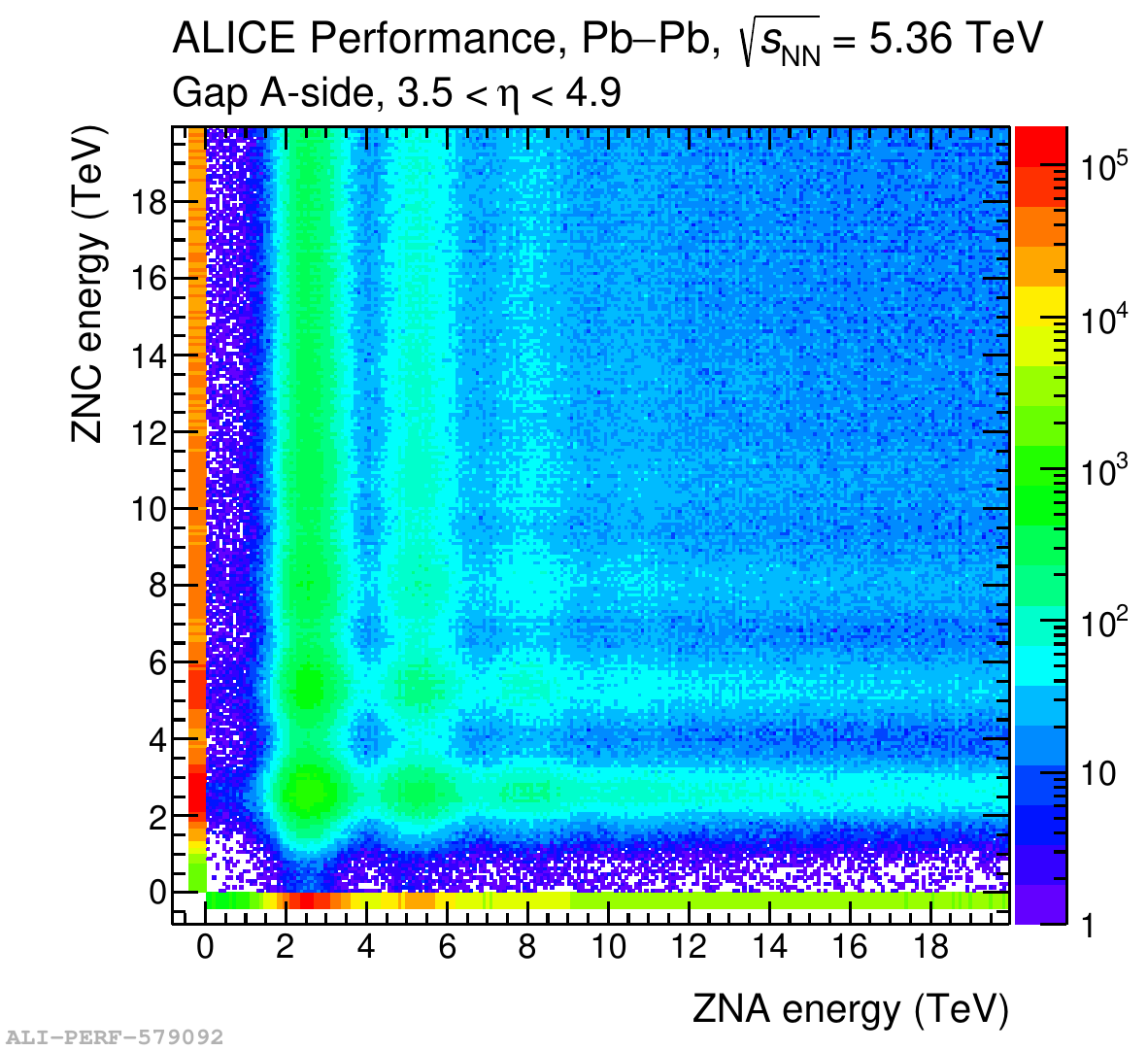}
  \end{center}
  \caption{Energy distribution in the neutron calorimeters on either side of the interaction point, in $+\eta$ gap events. The peaks in the distribution correspond to specific numbers of emitted neutrons. The 0nXn and Xn0n event classes predominantly fall into the underflow bins shown below \qty{0}{\TeV}.}\label{fig:zdc}
\end{figure}

ALICE is equipped with zero degree calorimeters (ZDCs), placed \qty{112.5}{\meter} along the beam direction on either side of the interaction point~\cite{Oppedisano:2009zz}. The neutron calorimeters, ZNA and ZNC, are used for detection of neutrons emitted from the colliding nuclei. The energy distributions measured by the ZNA and ZNC for events with a rapidity gap on the $+\eta$ side are shown in Figure~\ref{fig:zdc}. Peaks in the energy distributions coming from one or more neutrons can be identified. As expected, the energy distribution of the calorimeter sitting on the gap-side is dominated by low numbers of neutrons, with a steep fall-off towards higher values. On the other hand, the target-going side has a flatter distribution with a significant contribution from one or more neutrons. To further suppress contamination from hadronic collisions, events selected for further analysis are required to have no neutrons emitted on the photon-going side and at least one neutron emitted on the target going side, which we denote as 0nXn (Xn0n) for events with the photon-going side in the $+\eta$ ($-\eta$) direction.

$\PDzero$ mesons are reconstructed in the $\text{K}^-\pi^+$ decay channel in the midrapidity region ($|y|<0.9$). Tracking is done using the Time Projection Chamber (TPC)~\cite{Alme:2010ke, Lippmann:2014lay}, which provides three-dimensional space-point information for charged tracks. Kaons are identified by specific energy loss ($\text{d}E/\text{d}x$) in the TPC and velocity measured by the ALICE Time Of Flight (TOF) detector~\cite{ALICE:2000xcm}. Reconstruction of the $\PDzero$ meson decay vertex is performed using the ALICE Inner Tracking System (ITS)~\cite{ALICE:2013nwm}, allowing use of topological selections to improve the signal to background ratio.

\begin{figure}
    \centering
    \begin{subfigure}[b]{0.5\textwidth}
        \includegraphics[width=\linewidth]{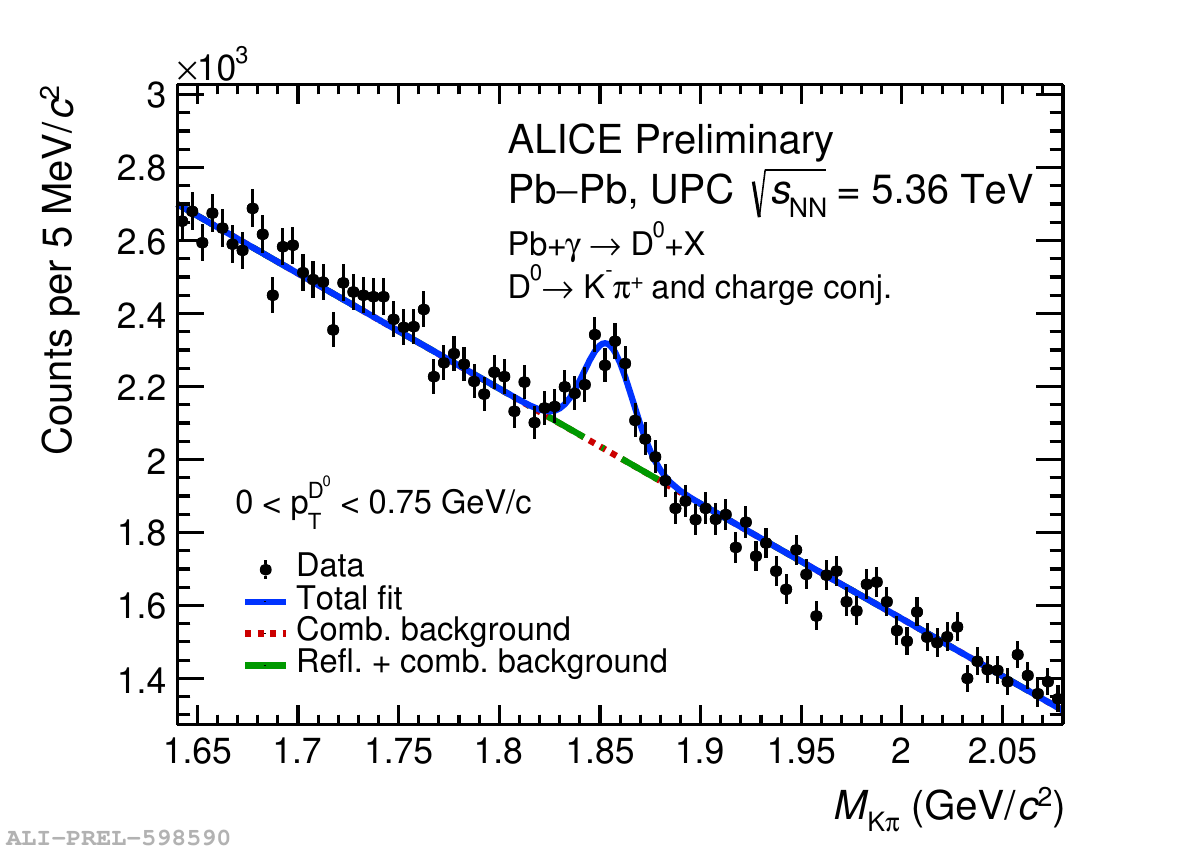}
    \end{subfigure}\hfill
    \begin{subfigure}[b]{0.5\textwidth}
        \includegraphics[width=\linewidth]{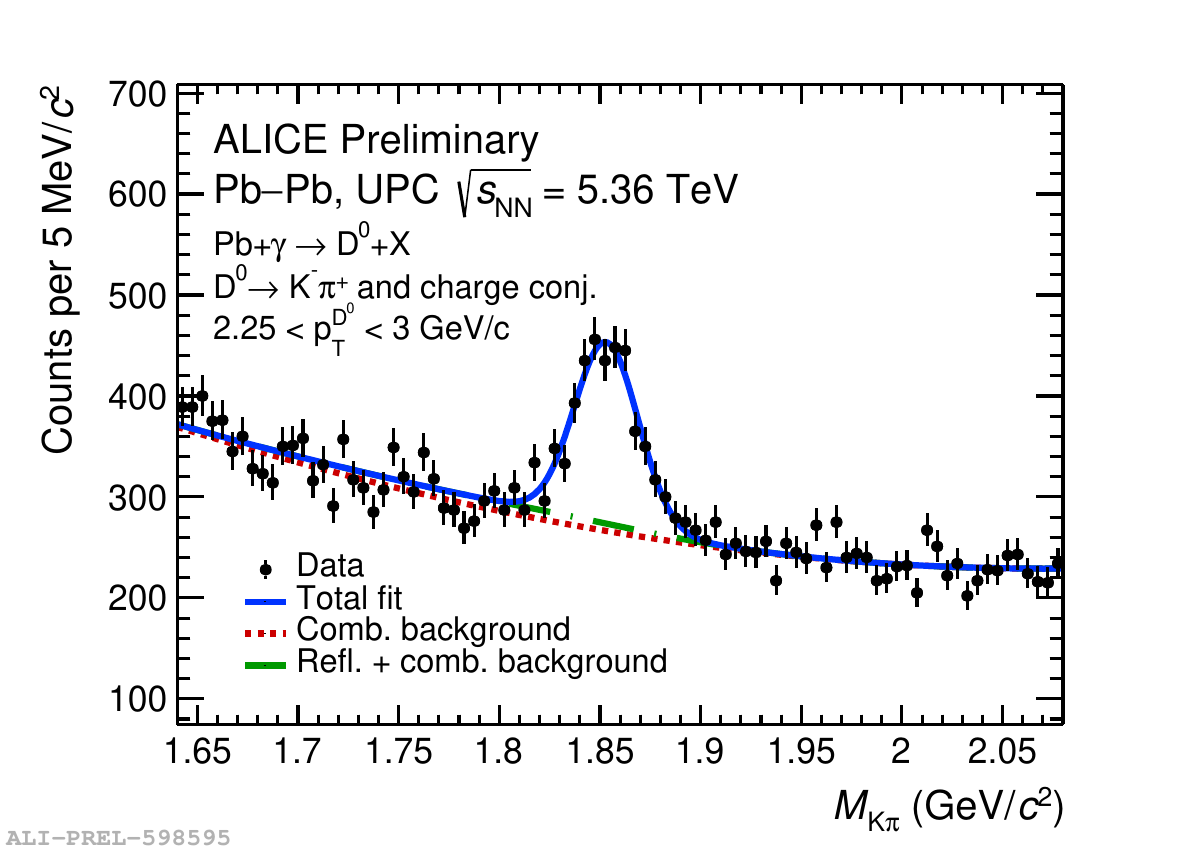}
    \end{subfigure}\hfill
    \caption{Invariant mass distributions of $\PDzero$ candidates in $-\eta$ gap events for two \pt bins.}
    \label{fig:yield}
\end{figure}

The yield is extracted in 8 bins of \pt from 0 to 12 GeV/$c$. Fits to the invariant mass yield in $-\eta$ gap events are shown for two of the bins in Figure~\ref{fig:yield}. In addition to the combinatorial background, there is a background contribution from pairs with swapped mass assumptions. The shape and relative normalization of this contribution is determined by simulations. The raw yield count of photoproduced $\PDzero$ mesons is corrected for acceptance and reconstruction efficiency using Monte-Carlo simulations. STARlight~\cite{Klein:2016yzr} is used to generate the photon spectrum, while DPMJET~\cite{Engel:1994vs} is used for the generation of photoproduced charm hadrons. 

\section{Results}
\label{sec-2}
\begin{figure}
    \centering
    \begin{subfigure}[b]{0.5\textwidth}
        \includegraphics[width=\linewidth]{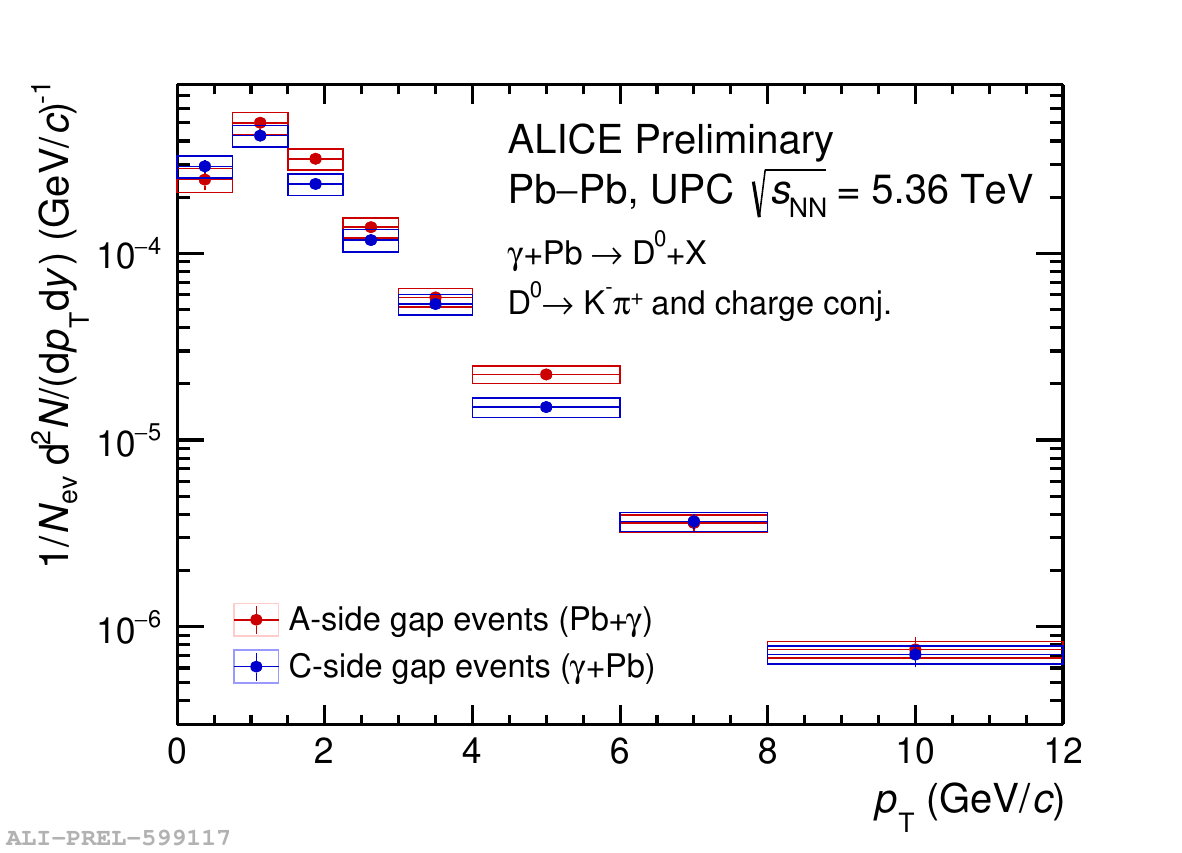}
    \end{subfigure}\hfill
    \begin{subfigure}[b]{0.5\textwidth}
        \includegraphics[width=\linewidth]{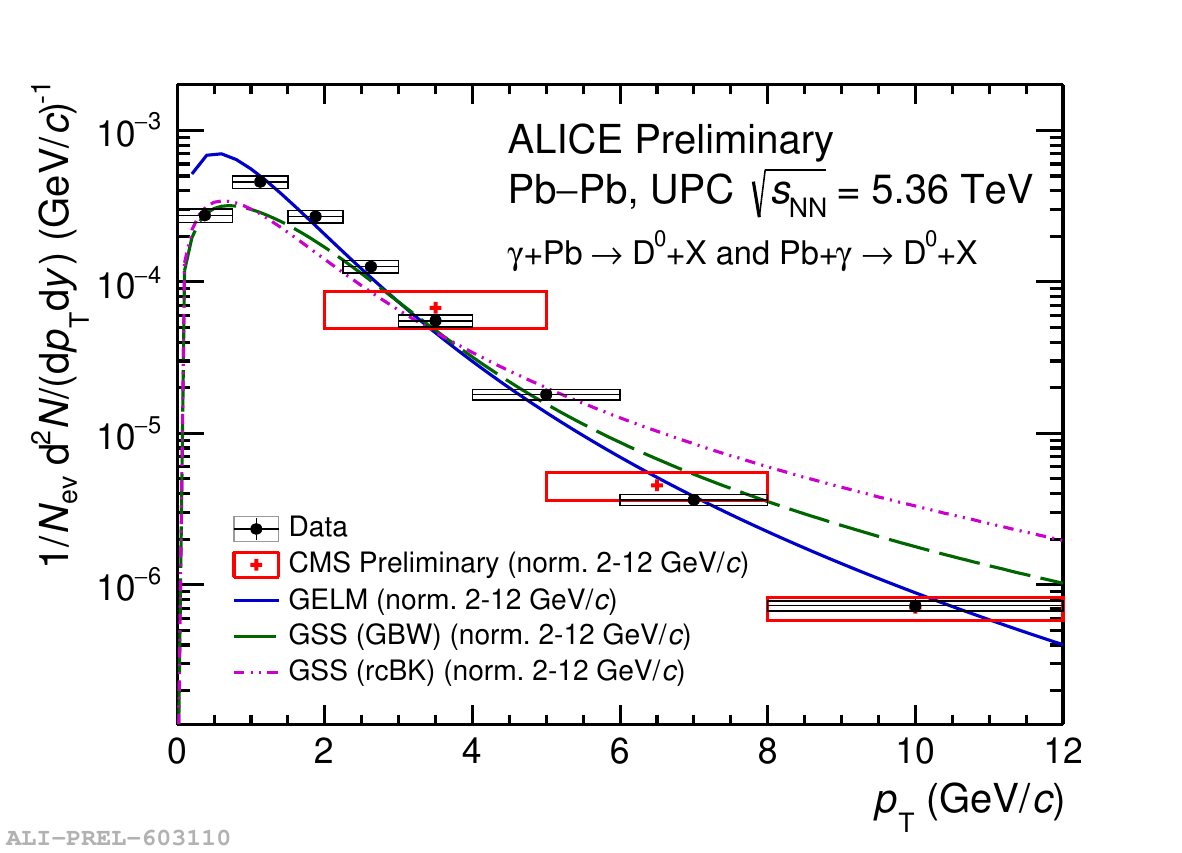}
    \end{subfigure}\hfill
    \caption{Left: Transverse momentum distribution of $\PDzero$ mesons in $+\eta$ side gap events and $-\eta$ gap events. Right: Combined transverse momentum distribution $\PDzero$ mesons from both event classes.}
    \label{fig:results}
\end{figure}

The corrected \pt distributions of photoproduced $\PDzero$ are shown in Figure~\ref{fig:results} (left), for $+\eta$ side and $-\eta$ gap events. The distributions are normalized by the number of visible events, as selected by the FIT and ZDC vetos described above. The two event classes correspond to slightly different rapidity gap selections, since the pseudorapidity coverage of the FT0A and FT0C detectors are not symmetric. However, we do not expect a significant dependence of the \pt distribution on the rapidity gap size. Therefore, the \pt distributions of the two event classes are combined, yielding a mean \pt of $1.476 \pm 0.026\,\text{(stat)}\pm 0.048\,\text{(syst.)}\; \text{GeV}/c$. The combined result is shown in Figure~\ref{fig:results} (right).

The \pt distribution, measured with high granularity down to $\pt=0$, is compared to several color glass condensate model calculations. Above $\pt \approx 2\;\text{GeV}/c$ the calculation by Gimeno-Estivill et al.~\cite{Gimeno-Estivill:2025rbw} gives a qualitatively good description of the shape of the \pt distribution, but the mean \pt is underestimated. Meanwhile, the two predictions by Gon\c{c}alves et al.~\cite{Goncalves:2017zdx} overestimate the mean \pt. The measured \pt shape is consistent with that of the preliminary cross section measurement by the CMS Collaboration~\cite{CMS:2024}. To facilitate comparisons of the shape of the \pt distribution, the curves and data points are normalized to the same integrated yield.

\section{Summary and outlook}
ALICE has measured the \pt distribution of photoproduced $\PDzero$ mesons from inelastic UPCs. The distribution was measured for the first time down to $\pt=0$, where the charm production is most sensitive to the low-$x$ gluon content of the target. Upcoming open charm photoproduction cross section measurements in ALICE will act as sensitive probes of the nuclear gluon distribution at low-$x$. Run 3 data is expected to allow measurement of several charm species in inelastic UPCs, allowing exploration of the system dependence of charm hadronization.

\end{document}